\documentclass[aps,pra,twocolumn,showpacs]{revtex4}

\usepackage{latexsym}

\usepackage{graphics}

\usepackage{psfig}

\begin{document}

\title{Optimal probabilistic estimation of quantum states}

\author{Jarom{\'\i}r Fiur\'{a}\v{s}ek}

\affiliation{Department of Optics, Palack\'y University,
     17.~listopadu 50, 772\,00 Olomouc, Czech~Republic}


\begin{abstract}

We extend the concept of probabilistic unambiguous discrimination of quantum states 
to quantum state estimation. We consider a scenario where the measurement device can output
either an estimate of the unknown input state or an inconclusive result. 
We present a general method how to evaluate the  maximum fidelity achievable by the
probabilistic estimation strategy.  We illustrate our method on two explicit
examples:  estimation of a qudit from a pair of conjugate qudits and phase
covariant estimation of a qubit from $N$ copies. We show that by allowing for 
inconclusive results it is possible to reach estimation fidelity higher than that
achievable by the best deterministic estimation strategy.

\end{abstract}


\maketitle

\section{Introduction}


The laws of quantum mechanics impose fundamental bounds on the amount of information
that can be extracted from the measurements on  quantum states. In particular, 
it is not possible to exactly determine an unknown quantum state from a single copy. 
More generally, nonorthogonal  quantum states cannot be perfectly deterministically 
discriminated  if only a single copy of the state is available. 
The optimal estimation and discrimination of quantum states has attracted a lot of
attention during recent years. This interest was largely stimulated by the
rapid development of quantum information theory.  Quantum measurement forms 
an essential ingredient of practically every protocol for
quantum information transmission and processing since it converts the quantum
information carried by a quantum system onto classical information.

Two different strategies were proposed in the literature to optimally discriminate 
among nonorthogonal quantum states.  The first option is to minimize the 
discrimination  error, i.e. the probability of a wrong guess
\cite{Holevo73,Yuen75,Helstrom76,Eldar01,Jezek02}. In this case the measurement device 
always provides a guess of the state, which sometimes may be wrong. 
An alternative approach  pioneered by
Ivanovic, Dieks and Peres \cite{Ivanovic87,Dieks88,Peres88} is the unambiguous probabilistic discrimination which 
allows for perfect error-free identification of non-orthogonal states at the expense of a
fraction of inconclusive results. It was shown that $n$ pure quantum states from a set
 $\{ |\psi_j\rangle\}_{j=1}^n$ can be unambiguously discriminated if and only if they are
linearly independent \cite{Chefles98,Chefles98b,Duan98,Zhang01,Sun01}. Recently, the concept of unambiguous discrimination was extended to mixed quantum
states \cite{Rudolph03,Raynal03,Eldar04,Feng04,Herzog04}. More generally, it was shown that by allowing for a fraction of inconclusive 
results it is possible to reduce the probability of a wrong guess  even if it is
not possible to achieve the perfect error-free discrimination \cite{Fiurasek03}.

The optimal quantum state estimation can be thought of as a limiting case of quantum
state discrimination among infinitely many states forming a continuous set. The
canonical example of the state estimation task is a determination of an unknown 
state of a qubit from a single copy if \emph{a-priori} the state could  lie anywhere
on the surface of the Bloch sphere \cite{Massar95}. The similarity of  the (mixed) 
estimated state $\rho_{\mathrm{est}}$ with the true pure state $|\psi\rangle$ is quantified by
the fidelity $F=\langle \psi|\rho_{\mathrm{est}}|\psi\rangle$ and the optimal state estimation
strategy is defined as the one which maximizes the  average estimation fidelity.
During recent years optimal estimation strategies were established for a wide class 
of input sets of states including optimal universal estimation of qubits
\cite{Massar95,Derka98} and qudits \cite{Bruss99},
optimal phase-covariant estimation of qubits \cite{Derka98} and qudits
\cite{Macchiavello03} and optimal estimation of coherent states \cite{Hammerer05}. 
Also optimal estimation of mixed states \cite{Vidal99} has been studied.

In this paper we generalize the concept of unambiguous state discrimination to state
estimation. We consider a scenario where the measuring apparatus can either output an
estimate of the state or an inconclusive result. We shall show on explicit examples
that with such probabilistic estimation strategy it is possible to increase the
estimation fidelity above that achievable by the optimal deterministic estimation.

The rest of the paper is structured as follows. In Section II we establish a general
formalism for the determination  of the optimal probabilistic covariant estimation
strategy. We shall show that the maximum attainable fidelity can be evaluated  as a
maximum eigenvalue of a certain operator. In Section III we apply the method to the
determination of optimal probabilistic estimation  of a qudit
state $|\psi\rangle$ from a pair of complex conjugate qudits $|\psi\rangle|\psi^\ast\rangle$.
This represents an extension of the well studied problem of optimal estimation of a
qubit from a pair of orthogonal qubits \cite{Gisin99,Massar00} to $d$-dimensional quantum system. 
Recently, the  optimal 
deterministic estimation of $|\psi \rangle$ from a single copy of 
$|\psi\rangle|\psi^\ast\rangle$ was addressed by Zhou et al. who numerically calculated
the maximum achievable fidelity \cite{Zhou06}. 
We shall show that this numerically obtained fidelity is actually the fidelity of
optimal probabilistic estimation and we will find a simple analytical formula for it. 
We will also derive from the first principles the maximum deterministic estimation 
fidelity $F_{\mathrm{det}}$ for this case. Remarkably, $F_{\mathrm{det}}$ turns out to coincide with 
the fidelity corresponding to the analytically  found local extremum point in Ref.
\cite{Zhou06}. 
As a second example we shall consider in Section IV the optimal phase covariant
estimation of a qubit from $N$ copies. We shall show that for $N>2$ the probabilistic
estimation strategy achieves strictly larger fidelity than the deterministic one and we
shall compare the asymptotic behavior of the fidelities for large $N$.
Finally, Section V contains conclusions and a brief summary of the main results.

\section{Optimal probabilistic estimation}

Consider a set of pure quantum states $|\Psi(\psi)\rangle$ which are 
parametrized by a state $|\psi\rangle$ and let $\mathcal{S}$ denote the set of all admissible
$|\psi\rangle$.  The a-priori probability distribution of $|\psi\rangle$ 
labeled by $d \psi$ satisfies $\int_{\mathcal{S}} d\psi=1$. 
Using this notation we can treat in a unified way more complex situations such as 
the estimation 
of $|\psi\rangle$ from $N$ copies of the state, when $|\Psi\rangle= |\psi\rangle^{\otimes N}$. 
 The goal of the quantum state estimation is to determine the state 
$|\psi\rangle$ as precisely as possible by performing a generalized quantum 
measurement described by a positive operator valued measure $\Pi(\phi) d\phi$ on
$|\Psi\rangle$. Here $|\phi\rangle \in \mathcal{S}$  and the
detection of $\Pi(\phi)$ implies that the state $|\phi\rangle$ is given as
the estimate. The optimal measurement strategy generally depends on 
the set $\mathcal{S}$ and on the  a-priori probability distribution $d \psi$. 

In  the probabilistic estimation one allows for inconclusive results where the machine
does not produce any estimate of the state. This null outcome is associated with a POVM
element $\Pi_0$ and the whole POVM should satisfy the completeness condition:
\begin{equation}
\int_{\mathcal{S}} \Pi(\phi) d \phi + \Pi_0 = \openone,
\label{POVMnormalization}
\end{equation}
where $\openone$ is the identity operator on the Hilbert space spanned by the states
$|\Psi(\psi)\rangle$. 
The success  of the estimation procedure can be conveniently quantified by the 
average fidelity. Consider first a particular input state $|\psi\rangle.$ The
normalized fidelity of the estimation of $|\psi\rangle$ can be expressed as
\begin{equation}
F(\psi)= \frac{1}{P(\psi)} \int_{\mathcal{S}} \langle
\Psi(\psi)|\Pi(\phi)|\Psi(\psi)\rangle |\langle \psi|\phi\rangle|^2 d \phi,
\label{Fpsi}
\end{equation}
where
\begin{equation}
P(\psi)= \int_{\mathcal{S}} \langle
\Psi(\psi)|\Pi(\phi)|\Psi(\psi)\rangle  d\phi
\label{Ppsi}
\end{equation}
is the probability of producing an estimate of the state and, consequently, 
$1-P(\psi)$ is the probability of inconclusive outcome.
We choose as a figure of merit the normalized average fidelity,
\begin{equation}
\bar{F}= \frac{\int_S F(\psi) P(\psi) d\psi}{\int_S P(\psi) d\psi}.
\label{F}
\end{equation}
The maximization of $\bar{F}$ is a complicated task since in general it requires  the
optimization of infinitely many POVM elements $\Pi(\phi)$.  The problem simplifies
 considerably  if the states $|\psi\rangle$ and $|\Psi(\psi)\rangle$ form orbits
of some group $G$,   $\mathcal{S}\equiv G$, and if $d \psi$ is an invariant measure induced by the Haar measure
on $G$. 
 In the rest of the paper we will assume that this is the case. 
We then have $|\psi\rangle= U(\psi)|0\rangle$ and
$|\Psi(\psi)\rangle=V(\psi)|\Psi(0)\rangle$, where $U(\psi)$ and
$V(\psi)$ denote unitary representations of the group $G$. With slight abuse of notation we
use $\psi$ to label the elements of $G$. It can be shown that due to the underlying
group structure the optimal POVM which maximizes $\bar{F}$ can always be chosen to be
covariant and all the POVM elements are generated from a single element,
\begin{equation}
\Pi_{C}(\phi)= V(\phi) \Pi_C V^\dagger(\phi),
\label{PiC}
\end{equation}
and $\Pi_C \equiv \Pi_C(0)$.

On inserting the expression (\ref{PiC}) into Eq. (\ref{F}) we obtain
\begin{equation}
\bar{F}= \frac{\mathrm{Tr}[R \Pi_C]}{\mathrm{Tr}[A \Pi_C]},
\label{Ftrace}
\end{equation}
where 
\begin{equation}
R= \int_{G} |\Psi(\psi)\rangle \langle \Psi(\psi)| 
\, |\langle \psi|0\rangle|^2 \, d \psi
\label{Rdef}
\end{equation}
and
\begin{equation}
A=\int_{G} |\Psi(\psi)\rangle \langle \Psi(\psi)| \, d \psi.
\label{Adef}
\end{equation}

Note that the expression (\ref{Ftrace}) is formally similar to the formula for the fidelity 
of the optimal probabilistic completely positive map  that approximates some 
unphysical operation which was derived in Ref. \cite{Fiurasek04}. Upon introducing 
\begin{equation}
\tilde{\Pi}_C= A^{1/2} \Pi_C A^{1/2}
\label{PiCtilde}
\end{equation}
we can rewrite Eq. (\ref{Ftrace}) as 
\begin{equation}
\bar{F} =\frac{\mathrm{Tr}[A^{-1/2}RA^{-1/2}\tilde{\Pi}_C]}{\mathrm{Tr}[\tilde{\Pi}_C]}.
\label{Fbar}
\end{equation}

It follows that the fidelity is bounded from above by the maximum eigenvalue 
$\mu_{\mathrm{max}}$ of the operator $M= A^{-1/2}RA^{-1/2}$. If the Hilbert space spanned by
$|\Psi(\psi)\rangle$ is finite dimensional then there exists a POVM which attains the
maximum fidelity $F_{\mathrm{max}}=\mu_{\mathrm{max}}$ and produces the 
estimate of $|\psi\rangle$ with nonzero probability $P > 0$.
Let  $|\mu_{j}^{\mathrm{max}}\rangle$, $j=1,\ldots, J$, be the
eigenvectors corresponding to the maximum eigenvalue $\mu_{\mathrm{max}}$. Then the 
POVM element $\Pi_{C,\mathrm{opt}}$ which generates the optimal covariant POVM 
can be expressed as
\begin{equation}
\Pi_{C,\mathrm{opt}}= A^{-1/2} \sum_{j,k=1}^J \pi_{jk} |\mu_j^{\mathrm{max}} \rangle 
\langle \mu_k^{\mathrm{max}}| A^{-1/2}.
\label{PiCopt}
\end{equation}
The coefficients $\pi_{jk}$ must be chosen such that $\Pi_{C,\mathrm{opt}} \geq 0$,
$\Pi_{C,\mathrm{opt}}^\dagger= \Pi_{C\mathrm{,opt}}$   and 
\begin{equation}
\int_{\mathcal{S}} V(\psi) \Pi_{C,\mathrm{opt}} V^\dagger(\psi) d \psi \leq \openone.
\label{PiCnormalization}
\end{equation}
The coefficients $\pi_{jk}$ may be optimized such as to maximize  the average probability of
success $P$ under the constraints $\Pi_{C,\mathrm{opt}} \geq 0$ and 
(\ref{PiCnormalization}). 
This is an instance of a semidefinite program which is a convex optimization problem that 
can be very efficiently  solved numerically \cite{Boyd96,Fiurasek04}. 

If the maximum eigenvalue is non-degenerate, $J=1$, then the optimal POVM element 
$\Pi_{C,\mathrm{opt}}$ is proportional to a rank-one projector
\begin{equation}
\Pi_{C,\mathrm{opt}}= \frac{1}{\mathcal{N}} A^{-1/2} |\mu^{\mathrm{max}} \rangle \langle
\mu^{\mathrm{max}}| A^{-1/2}
\label{PiCoptrankone}
\end{equation}
and the normalization constant $\mathcal{N}$ has to be chosen such that 
(\ref{PiCnormalization}) holds.

\section{Pair of conjugate qudits}

\subsection{Optimal probabilistic estimation}

In this section we will investigate the optimal probabilistic estimation of a state of
a single qudit $|\psi\rangle$ from a pair of conjugate qudits,
$|\Psi(\psi)\rangle=|\psi\rangle|\psi^\ast\rangle$. The a-priori distribution $d \psi$
is assumed to be induced by the Haar measure on the group $\mathrm{SU}(d)$. 
This scenario is a generalization
of the estimation of a qubit from a pair of orthogonal qubits  \cite{Gisin99,Massar00} 
to dimensions  $d > 2$. The corresponding operators $A$ and $R$  can be easily evaluated 
with the help of the Schur Lemma. The unitary representation $U^{\otimes N}$ of the group 
$\mathrm{SU}(d)$  acts irreducibly on the totally symmetric subspace 
of $N$  qudits. Taking into account that $|\psi^{\ast}\rangle\langle \psi^{\ast}|=
(|\psi\rangle\langle \psi|)^T$ and exchanging the order of integration and (partial)
transposition we  obtain
\begin{equation}
A=\frac{1}{d(d+1)} (\openone+d \Phi^{+}),
\label{Aqudit}
\end{equation}
where $\Phi^+=|\Phi^{+}\rangle\langle \Phi^{+}|$ and
\begin{equation}
|\Phi^{+}\rangle=\frac{1}{\sqrt{d}} \sum_{j=0}^{d-1} |j\rangle|j\rangle
\label{Phiplus}
\end{equation}
is a maximally entangled state of two qudits. We also find that
\begin{equation}
R=\frac{1}{D_3^{+}(d)}\mathrm{Tr}_{3}[\openone_1\otimes \openone_{2}\otimes
|0\rangle_3\langle 0| \, (\Pi_{123}^{+})^{T_2}].
\label{Rquditdef}
\end{equation}
Here $\Pi_{123}^{+}$ is the projector onto the symmetric subspace of three qudits,
 $D_3^+(d)=\frac{1}{6}d(d+1)(d+2)$  is the dimension of this
subspace, $T_2$ denotes partial transposition with respect to the second qudit, 
and $\mathrm{Tr}_3$ stands for the partial trace over the third qudit. 
After some algebra we find  that $R$ can be expressed as
\begin{eqnarray}
R&=&\frac{1}{d(d+1)(d+2)} \left[\openone_{12}+d \Phi^{+}
+\sqrt{d}\,|\Phi^{+}\rangle\langle00| \right .\nonumber \\
&& \left.+\sqrt{d}\,|00\rangle\langle\Phi^{+}|+ \openone_1\otimes |0\rangle_2\langle 0| +
 |0\rangle_1\langle 0|\otimes \openone_2\right].
 \nonumber \\
 \label{Rqudit}
\end{eqnarray}

Instead of calculating the eigenvalues of $A^{-1/2}R A^{-1/2}$ we can equivalently look
for the eigenvalues of $RA^{-1}$ because the eigenvalues of these two operators coincide
and the latter is easier to deal with. The operator $A$ can be easily inverted, 
\begin{equation}
A^{-1}= d(d+1)\left[ \openone - \frac{d}{d+1}\Phi^{+}\right]
\label{Aquditinv}
\end{equation}
and we arrive at
\begin{eqnarray}
RA^{-1} &=& \frac{1}{d+2}\left[\openone_{12}-\frac{d}{d+1} \Phi^{+}
+\sqrt{d}\,|\Phi^{+}\rangle\langle 00| \right. \nonumber \\
&& \left.-\frac{\sqrt{d}}{d+1}\,|00\rangle\langle\Phi^{+}|+ \openone_1\otimes |0\rangle_2\langle 0| +
|0\rangle_1\langle 0|\otimes \openone_2\right].
\nonumber \\
\label{RAinv}
\end{eqnarray}

This operator possesses only four different eigenvalues which can be expressed
analytically for arbitrary $d$. The eigenvalue
$\mu_1=1/(d+2)$ is $d(d-2)$-fold degenerate and the eigenstates read 
$|j\rangle_1|k\rangle_2$, where $j\neq 0,$ $k \neq 0$. The second eigenvalue
$\mu_4=2/(d+2)$ is $(2d-2)$-fold degenerate with eigenstates $|0\rangle_1|j\rangle_2$
and  $|j\rangle|_10\rangle_2$, $j \neq 0$. Finally, the last two eigenvalues are
non-degenerate and can be expressed as
\begin{equation}
\mu_{3,4}= \frac{2}{d+2}\left[1\pm\sqrt{\frac{d}{2(d+1)}}\right].
\label{mu34}
\end{equation}
The maximum eigenvalue  is $\mu_3$ for all $d \geq 2$ and the fidelity of the 
optimal probabilistic estimation of $|\psi\rangle$ form a pair of conjugate qudits
is equal to this eigenvalue,
\begin{equation}
F_{\mathrm{max,prob}}=\frac{2}{d+2}\left[1+\sqrt{\frac{d}{2(d+1)}}\right].
\label{Fquditprobmax}
\end{equation}
The numerical values for the optimal estimation fidelity $F_{\perp}$ obtained by Zhou
\emph{et al.} \cite{Zhou06} fully agree with the above analytical formula so 
their global optimization actually yielded the optimal probabilistic estimation 
strategy. 

The optimal probabilistic covariant POVM  is generated by the POVM element 
$\Pi_{C,\mathrm{opt}}=|\pi_{C,\mathrm{opt}}\rangle \langle \pi_{C,\mathrm{opt}}|,$ where
\begin{equation}
|\pi_{C,\mathrm{opt}}\rangle \propto
 |0 0\rangle -\sqrt{\frac{2d}{d+1}}\frac{\sqrt{2(d+1)}-\sqrt{d}}{d+2}\,|\Phi^{+}\rangle.
 \label{piprobopt}
\end{equation}

\subsection{Optimal deterministic estimation strategy}

For $d>2$ the optimal estimation strategy  obtained above cannot be made 
deterministic and there is a nonzero probability of inconclusive results.
Thus a question arises what is the optimal deterministic strategy of estimation of 
$|\psi\rangle$ from a single copy of the state $|\psi\psi^\ast\rangle$.
When seeking an answer to this question we can restrict ourselves to the covariant
POVMs. Since the probability of inconclusive results should vanish, we have $\Pi_0=0$
and the completeness condition for the POVM becomes
\begin{equation}
\int_S U(\psi) \otimes U^{\ast}(\psi) \Pi_C U^\dagger (\psi) \otimes U^T(\psi) d \psi
= \openone.
\label{PiCquditnormalization}
\end{equation} 
Recall that $U(\psi)$ is a unitary acting on the Hilbert space of  a single qudit and
$U(\psi)|0\rangle=|\psi\rangle$.
We should maximize the estimation fidelity $F=\mathrm{Tr}[\Pi_C R]$ 
under the above completeness condition. In the present case this is equivalent 
to maximizing $F$ under simpler constraints that can be obtained 
from (\ref{PiCquditnormalization}). In particular, by calculating the 
trace of Eq. (\ref{PiCquditnormalization}) and by 
taking into account the invariance $U\otimes U^\ast |\Phi^{+}\rangle=|\Phi^{+}\rangle$
 we find that 
\begin{equation}
\mathrm{Tr}[\Pi_C]=d^2, \qquad \mathrm{Tr}[\Pi_C \Phi^{+}]=1
\label{Pi0constraints}
\end{equation}
must hold.

The constraints (\ref{Pi0constraints}) can be accounted for by introducing 
two Lagrange multipliers $\lambda_1$ and $\lambda_2$ and our task is to maximize
\begin{equation}
\mathcal{F}[\Pi_C]=\mathrm{Tr}[R \Pi_C] -\lambda_1 \mathrm{Tr}[\Pi_C]-\lambda_2
\mathrm{Tr}[\Phi^{+}\Pi_C]
\label{Fcalligraphic}
\end{equation}
under the constraints (\ref{Pi0constraints}) and $\Pi_C \geq 0$. 
This is an instance of a semidefinite program \cite{Boyd96}. 
For this class of convex optimization problems one can straightforwardly 
derive the extremal equation for the optimal $\Pi_C$  and we obtain
\begin{equation}
(R-\lambda_1 \openone-\lambda_2 \Phi^{+}) \Pi_C =0.
\label{PiCextremal}
\end{equation}
Moreover, we also find the optimality condition,
\begin{equation}
\lambda_1 \openone + \lambda_2 \Phi^{+} -R \geq 0. 
\label{Optimality}
\end{equation}
If (\ref{PiCextremal}) and (\ref{Optimality}) hold simultaneously, then $\Pi_C$ is 
the optimal one which maximizes the fidelity. To prove this statement we 
take the trace of a product of Eq. (\ref{Optimality})
with an arbitrary $\tilde{\Pi}_C$ which satisfies all the constraints imposed on it. 
We get $\mathrm{Tr}[\tilde{\Pi}_C R] \leq \lambda_1 d^2+\lambda_2$ hence the Lagrange
multipliers provide an upper bound on the achievable fidelity which is saturated 
if the POVM satisfies Eq. (\ref{PiCextremal}).
 
The optimal POVM has qualitatively similar structure as the optimal probabilistic POVM
(\ref{piprobopt}), namely, 
$\Pi_{C,\mathrm{det}}= |\pi_{C,\mathrm{det}}\rangle\langle \pi_{C,\mathrm{det}}|$ where
\begin{equation}
|\pi_{C,\mathrm{det}}\rangle= 
\sqrt{d(d+1)} |00\rangle - (\sqrt{d+1}-1)|\Phi^{+}\rangle.
\label{pi0det}
\end{equation}
By construction, this POVM  satisfies the completeness 
condition (\ref{PiCquditnormalization}). On inserting $\Pi_{C,\mathrm{det}}$ into 
Eq. (\ref{PiCextremal}) we can solve for the Lagrange multipliers and we get
\begin{eqnarray}
\lambda_1&=&\frac{4-(1-\sqrt{\frac{1}{d+1}})(1+\frac{2}{d})}{d(d+1)(d+2)},
\nonumber \\
\lambda_2&=& \frac{1}{d(d+1)(d+2)} 
\left[\frac{d^3+2d^2-2d-4}{d\sqrt{d+1}}+\frac{4}{d}+d\right]. \nonumber \\
\label{lambda12}
\end{eqnarray}
This choice guarantees that Eq. (\ref{PiCextremal}) holds for any $d$. To prove 
the optimality of the POVM (\ref{pi0det}) it remains to check the inequality 
(\ref{Optimality}). Since $\lambda_1 > 2/[d(d+1)(d+2)]$  the only nontrivial part is
the  verification of the positive semidefiniteness of the operator 
in the two-dimensional subspace spanned by  $|00\rangle,~|\Phi^{+}\rangle$. 
Let $\Pi_2$ denote the projector onto this subspace and consider the $2 \times 2 $ matrix 
$K= \Pi_{2}(\lambda_1 \openone + \lambda_2 \Phi^{+} -R)\Pi_2$.
One eigenvalue of  $K$ is zero due to the optimality condition (\ref{PiCextremal}). To prove that 
$ K \geq 0 $ it thus suffices to show that $\mathrm{Tr} K \geq 0$ and after some algebra
we find 
  \begin{equation}
 \mathrm{Tr} K = 2 \lambda_1+\lambda_2 -\frac{d+6}{d(d+1)(d+2)}.
 \label{TrM}
 \end{equation}
It can be shown that $\mathrm{Tr}K$ is a growing function of $d$  and that it is
positive for all integer $d \geq 2$. This concludes the optimality proof.
 
The fidelity of the optimal deterministic estimation corresponding to the optimal
covariant POVM reads
\begin{equation}
F_{\mathrm{max,det}}=\frac{1}{d^2(d+2)} \left[ 3d^2-4d+4+\frac{2d^2+2d-4}{\sqrt{d+1}}\right].
\end{equation} 
This expression  agrees with the formula given by Zhou \emph{et al.} \cite{Zhou06}. 
In that paper, the authors claimed that this is only a local maximum of the fidelity 
and they calculated the global maximum of the fidelity numerically. 
Our findings provide a precise
interpretation of their results. The local maximum \emph{is} in fact the maximum
achievable fidelity of deterministic estimation from a pair of conjugate qudits while
the global maximum given in Ref. \cite{Zhou06} corresponds to the optimal probabilistic estimation
strategy which allows for inconclusive results.

It is instructive to explicitly evaluate the probability that the machine outputs 
an estimate $|0\rangle$ for an input state $|\psi\psi^\ast\rangle$. We have 
\begin{equation}
P_{\perp}(0|\psi)= \frac{1}{d}\left|d\sqrt{d+1}|\langle \psi|0\rangle|^2-\sqrt{d+1}+1 \right|^2 .
\label{Pperp}
\end{equation}
Note that this probability is zero if the overlap of the true state $|\psi\rangle$ 
with the estimated state $|0\rangle$ is equal to $|\langle \psi
|0\rangle|^2=(\sqrt{d+1}-1)/(d\sqrt{d+1})$. It is interesting to compare this with the 
optimal estimation from a pair of identical qudits $|\psi \psi\rangle$, where the
optimal covariant POVM is generated by  $|\pi_{C,||}\rangle=\sqrt{d(d+1)/2}|00\rangle$  
and the corresponding probability of guessing $|0\rangle$ for the input state 
$|\psi\rangle$ reads $P_{||}(0|\psi)= \frac{1}{2}d(d+1)|\langle \psi|0\rangle|^4$
which vanishes only if the state $|\psi\rangle$ is orthogonal to $|0\rangle$.
In particular, for $d=2$ the probability $P_{||}=0$ only if $|\psi\rangle=|1\rangle$ 
while $P_\perp$ is zero for all states on a certain circle of the Bloch sphere.
This observation gives some more insight into why the state $|\psi\rangle$ can be  
estimated with higher precision from $|\psi \psi^\ast\rangle$ than from $|\psi
\psi\rangle$.

\section{Optimal probabilistic estimation of equatorial qubits}

In this section we will investigate the optimal phase-covariant 
probabilistic estimation of a qubit.  We shall assume that it is \emph{a-priori} known that 
the qubit state is located on the equator of the Poincare sphere,
$|\psi(\varphi)\rangle=\frac{1}{\sqrt{2}}(|0\rangle + e^{i\varphi} |1\rangle).$
The state is thus characterized by a single parameter - the relative phase $\varphi$.
Starting from the seed state $\frac{1}{\sqrt{2}}(|0\rangle+|1\rangle)$ all 
states $|\psi(\varphi)\rangle$ can be obtained as the orbit of the Abelian group 
$\mathrm{U}(1)$  which generates rotations of the Bloch sphere about $z$ axis. 
To make our treatment general we will consider optimal estimation 
of $|\psi(\varphi)\rangle$ from $N$ input copies.

This scenario corresponds to a typical phase-shift measurement, where 
$N$ particles pass through an interferometer  which applies an unknown relative 
phase shift $\varphi$ to one of the states of the particle and the goal is 
to determine $\varphi$ as precisely as possible.  The optimal phase-estimation strategies 
which reach the so-called Heisenberg limit $\Delta \varphi \approx \frac{1}{N}$ require 
entangled input states of $N$ particles \cite{Luis00}. 
Here we show that even for the product state
$|\psi(\varphi)\rangle^{\otimes N}$ it is possible to probabilistically 
improve the precision of $\varphi$ estimation, as witnessed by the improved asymptotic scaling
of the optimal fidelity $1- F_{\mathrm{max,prob}} \propto \frac{1}{N^2}$. However, it should be noted that this
apparent improvement is achieved only for the sub-ensemble of conclusive measurement
outcomes while the inconclusive outcomes are neglected.

The input state  $|\psi(\varphi)\rangle^{\otimes N}$ belongs to the $N+1$ dimensional 
symmetric (bosonic)  subspace of the Hilbert space of $N$ qubits and it can be written
as follows,
\begin{equation}
|\psi(\phi)\rangle^{\otimes N} = \frac{1}{2^{N/2}} \sum_{k=0}^N \sqrt{N \choose k}
e^{ik\varphi} |N;k\rangle.
\end{equation}
Here $|N;k\rangle$ denotes a normalized fully symmetric state of $N$ qubits with 
$k$ qubits in state $|1\rangle$ and $N-k$ qubits in state $|0\rangle$.

The operators $A$ and $R$ can be determined from the formulas (\ref{Adef}) and
(\ref{Rdef}), where we have
to integrate over the phase shift $\varphi$, $\int d\psi=\int_0^{2\pi} \frac{1}{2\pi}
d\varphi$. After the integration we obtain
 \begin{equation}
 A= \frac{1}{2^N} \sum_{k=0}^N {N \choose k} |N;k\rangle \langle N;k|
 \label{Aequatorial}
 \end{equation}
and 
 \begin{eqnarray}
 R&=& \frac{1}{2^{N+1}}\sum_{k=0}^N {N \choose k} |N;k\rangle \langle N;k|
\nonumber \\
 &&+\frac{1}{2^{N+2}} \sum_{k=1}^N \sqrt{{N \choose k}{N \choose k-1}} 
 (X_k+X_k^\dagger),
 \label{Requatorial}
 \end{eqnarray}
 where $X_k=|N;k\rangle\langle N;k-1|$.
Since the operator $A$ is diagonal in the basis $|N;k\rangle$, the operator $M$ whose
maximum eigenvalue determines the maximum achievable estimation fidelity can be easily
calculated and we have
\begin{equation}
M=A^{-1/2} R A^{-1/2}= \frac{1}{2} \openone 
+\frac{1}{4} \sum_{k=1}^N (X_k+X_k^\dagger).
\label{Mequatorial}
\end{equation}

\begin{figure}[!t!]
\centerline{\psfig{figure=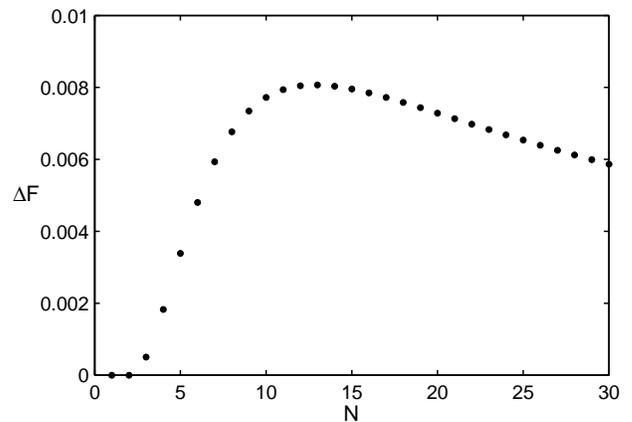,width=0.95\linewidth}}
\caption{The difference of the fidelities of optimal probabilistic and deterministic 
estimation $\Delta F=F_{\mathrm{max,prob}}-F_{\mathrm{max,det}}$ is plotted 
as a function of the number of copies $N$ of the qubit state $|\psi(\varphi)\rangle$.}
\end{figure}

Instead of directly working with $M$ let us consider the operator 
$\tilde{M}=4M-2 \openone$ and the eigenvalues $\mu_j$ of $M$ are then related to the 
eigenvalues $\tilde{\mu}_j$ of $\tilde{M}$ by $\mu_j=(\tilde{\mu}_j+2)/4$. 
The matrix $\tilde{M}$ is tridiagonal and its characteristic  polynomial 
is given by the Tchebychev polynomial of the second kind,
$\mathrm{det}(\tilde{M}-\lambda \openone)=U_{N+1}(-\frac{\lambda}{2})$. 
Maximum eigenvalue of $\tilde{M}$ is thus given by  the largest root of $U_N$ and 
with the help of the definition $U_N(\cos\theta)=\sin[(N+1)\theta]/\sin\theta$ 
we arrive at \cite{Bagan04}
\begin{equation}
\tilde{\mu}_{\mathrm{max}} = 2 \cos \left(\frac{\pi}{N+2} \right).
\label{mutildemax}
\end{equation}
The maximum fidelity can be determined as $(\tilde{\mu}_{\mathrm{max}}+2)/4$
and  we finally obtain
\begin{equation}
F_{\mathrm{max,prob}}= \frac{1}{2}\left[ 1 + \cos \left(\frac{\pi}{N+2} \right)
\right].
\label{Fmaxequatorial}
\end{equation}

It is instructive to compare the fidelity $F_{\mathrm{max,prob}}$ with the fidelity of
the optimal deterministic phase covariant estimation of a qubit from $N$ copies
\cite{Derka98},
\begin{equation}
F_{\mathrm{max,det}}=\frac{1}{2}+
\frac{1}{2^{N+1}}\sum_{k=1}^N \sqrt{{N \choose k}{N \choose k-1}}.
\label{Fmaxdetequatorial}
\end{equation}
It follows that for $N=1$ and $N=2$  $F_{\mathrm{max,prob}}=F_{\mathrm{max,det}}$ hence
it is not possible to improve the fidelity of estimation by allowing for some fraction
of inconclusive results. However, if $N \geq 3$ then 
$F_{\mathrm{max,prob}}>F_{\mathrm{max,det}}$ and the optimal probabilistic estimation strategy
attains a strictly larger fidelity than the optimal deterministic strategy. This is
illustrated in Fig. 1 which shows the difference between the fidelities 
(\ref{Fmaxequatorial}) and (\ref{Fmaxdetequatorial}).

\begin{figure}[!t!]
\centerline{\psfig{figure=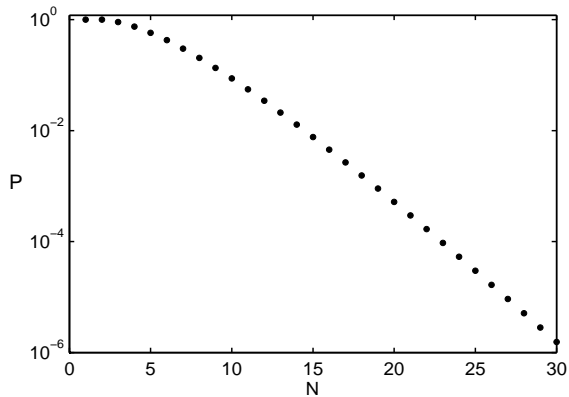,width=0.9\linewidth}}
\caption{Probability of successful estimation $P$ versus the number of copies $N$
of the qubit state $|\psi(\varphi)\rangle$.}
\end{figure}

Since the maximum eigenvalue of $M$ is non-degenerate, the
optimal covariant POVM is generated by the rank-one projector (\ref{PiCoptrankone}). 
The maximization of
the success probability $P=\mathrm{Tr}[\Pi_{C,\mathrm{opt}}A]$ is equivalent to the minimization
of $\mathcal{N}$ under the constraint  (\ref{PiCnormalization}). Let 
$|\mu_{\mathrm{max}}\rangle=\sum_{k=0}^N c_k|N;k\rangle$ be the normalized eigenvector
of $M$ with the  eigenvalue $\mu_{\mathrm{max}}$. Then it is optimal to choose
\begin{equation}
\mathcal{N}= 2^N \max_k {N \choose k}^{-1}|c_k|^2. 
\end{equation}
Numerical calculation reveals that  $P$ decreases 
exponentially with growing $N$, see Fig. 2.

The further facilitate the comparison of the  fidelities  let us analyze 
their asymptotic behavior for large $N$. For the
probabilistic estimation strategy we obtain
\begin{equation}
F_{\mathrm{max,prob}} \approx 1-\frac{\pi^2}{4} \frac{1}{(N+2)^2}
\end{equation}
and we can write $1-F_{\mathrm{max,prob}} =O(N^{-2})$. On the other hand, for the
deterministic estimation we find that $1-F_{\mathrm{max,det}} \approx O(N^{-1})$. 
We can see
that with growing $N$ $F_{\mathrm{max,prob}}$ converges to unity much faster than 
$F_{\mathrm{max,det}}$. This superior scaling is achieved at the expense of 
a decreasing  probability of successful estimation $P$.

\section{Conclusions}

In the present paper we have generalized the concept of unambiguous quantum state
discrimination to the realm of quantum state estimation. We have shown that by allowing
for inconclusive results  it is possible to increase the fidelity of estimation evaluated
for the sub-ensemble of conclusive outcomes of the estimation process. We have
established a general formula for the maximum fidelity achievable by  the probabilistic
state estimation.  The method was illustrated on two explicit examples. First, we have studied
the optimal estimation of a qudit from a pair of conjugate qudits and we have provided
an exact interpretation of the results recently obtained by Zhou \emph{et al.}
\cite{Zhou06}. As a second example we have investigated the phase covariant estimation
of a qubit from $N$ copies of the state.  The  present quantum-state estimation scheme 
could find applications in quantum communication and it may potentially 
help to probabilistically 
 improve the sensitivity and precision of measurements  performed at the quantum
limit.

\begin{acknowledgments}

This research was supported by the projects MSM6198959213 and
LC06007 of the Ministry of Education of the Czech Republic.

\end{acknowledgments}

\end{document}